
\input phyzzx
\frontpagetrue
\title{ Perturbative analysis of
an n-Ising model on a random surface }
\author{ Shinobu Hikami$^{1}$ and Edouard Br\'ezin$^{2}$ }
\address{$^{1}$ Department of Pure and Applied Sciences,
University of Tokyo, Meguro-ku, Komaba, 3-8-1, Tokyo 153,
e-mail address:a87353@tansei.cc.u-Tokyo.ac.jp}
\address{$^{2}$Laboratoire de Physique Th\'eorique de l'
Ecole Normale Sup\'erieure,24 rue Lhomond,F-75231 Paris
Cedex 05, e-mail address:brezin@physique.ens.fr
\foot{Unit\'e prope du Centre National de la Recherche
Scientifique,associ\'ee ${\grave a}$ l'Ecole Normale Sup\'erieure
et ${\grave a}$ l'Universit\'e Paris-Sud.} }
\abstract{ Two dimensional quantum gravity coupled to a conformally
invariant matter field of central charge $c=n/2$, is represented, in
a discretized version, by $n$ independent Ising spins
per cell of the triangulations of a random surface.
The matrix integral representation of this model leads to a
diagrammatic expansion in powers of the cosmological constant
for fixed genus. From the behaviour of this expansion at large orders,
when the Ising coupling constant is tuned to criticality,
one extracts the values of the string susceptibility exponent.
We extend our previous calculation to order eight for genus zero
and investigate now also the genus one case in order to
check the possibility of having a well-defined double scaling
limit even for $c>1$.
\break
               \hskip 10cm September 1992 }
\par
  A discretized approach to two-dimensional quantum gravity
coupled to matter fields of central charge $c=n/2$, is
easily formulated as an integral over $2^{n}$ $N\times N$ Hermitian
matrices\ref{1-3}. In this model each cell of a discretized random
surface carries $n$ Ising spins.
For a fixed triangulation these $n$ species do not interact and the
matter partition function is simply the $n$-th power of one Ising spin;
however the various species are effectively coupled by
the sum over the triangulations. In a previous article\ref{1},
we found that at relatively small order in perturbation
theory in the cosmological constant, we recovered with a
surprizing accuracy, the known results for the string
susceptibility exponent $\gamma_{string}$, for $c<1$ and
planar surfaces.
For larger values of $c$, at least up to $c=4$, we did not
see any sign of the tachyonic instability of the theory.
Similar conclusions on the absence of instabilities were
obtained through Monte-Carlo simulations of the same model\ref{2,3},
but the values of the exponents were not reported. There
were several attempts to evaluate the string susceptibility
exponent for $c>1$, by generating random surfaces embedded in a
space of more than one dimension, but the results have not
really converged so far\ref{4-7}.
\par
  In our previous work\ref{1}, we had expanded the free energy up to sixth
order in the cosmological constant $g$, for surfaces of genus zero.
In this letter we extend the series to order eight, and include
surfaces of genus one, in order to check the variation of the
string susceptibility with genus which governs the existence of
a double scaling limit.  The expansion in powers of $g$ is given
by $\phi^{4}$ Feynman diagrams in which each vertex carries $n$
spins, i.e. $2^{n}$ "flavors". The Ising interaction $\beta$ is present
in the relative weight of the up-up and up-down propagators\ref{7}
through the parameter $a$ defined as

$$     a = \exp ( - 2\beta )  \eqno{(1)} $$

The sum over the $n$-Ising species is, for a given diagram, the
$n$-th power of a one Ising sum, which is a polynomial in the
variable $a^{2}$. When the parameter $a$ varies and reaches its
critical value, for instance 1/4 for one Ising spin, the theory
crosses over from pure gravity, with $\gamma_{string}=-1/2$ to
a coupling with conformally invariant matter with central charge
$n/2$ and an exponent $\gamma_{string}$, which is for one Ising
spin, but that we have tried to determine for $n$ spins.
\par
The vertex of the $\phi^{4}$ hermitian matrix model, carries
four different lines of different colors $i$, $i=1,...,N$.
At $l$-th order the permutations of the $4l$ color indices,
generate the various diagrams, and this requires a fast algorithm.
If we use unitary instead of hermitian matrices, we obtain the same
type of diagrams, with two incoming and two outgoing lines
at each vertex, and the diagrams differ only by symmetry factors.
In fact these symmetry factors are the same for the two ensembles
at genus zero, but there are differences at genus one.
With this unitary ensemble we generate the Feynman diagrams with
the same tree sorting algorithm that was used for the study of
the perturbation theory at large order of the Landau-Ginzburg
Hamiltonian\ref{8}.  In this method, the symmetry factor is
automatically taken acount by the generation of all possible
line combinations. Disconnected diagrams are avoided by a computation
of the determinant of the adjacency matrix; it is made of 2 on the diagonal
, and -1 for the elements corresponding to vertices
connected by a line and some of the -1 of the element is replaced by 0.
This determinant is equal to the number of Euler trails for a connected
graph, but it vanishes for a disconnected  diagram.\ref{8}
\par
We have a certain number of checks on our eighth order result: the one
Ising spin first\ref{9}. Then, if either $a$ or $n$ vanish we must recover the
series of pure gravity\ref{10}.
The free energy is given by
$$ {F\over{2^{n}}}=\sum_{k} C_{k}g^{k} \eqno{(2)}$$
and the coefficient $C_{k}$ behaves for the large order as
$$ C_{k} \simeq A^{k}k^{-3+\gamma_{string}} \eqno{(3)}$$
The ratio $f_{k}=C_{k}/C_{k-1}$ gives the  value of
the string susceptibility exponent by the extrapolation, for example
we have used the following ratio method in the previous
analysis.\ref{1}

$$ A_{k}=(1+k)f_{k}-kf_{k-1} \eqno{(4)}$$
$$ \gamma_{string}=3-k(1+k)(f_{k}-f_{k-1})/A_{k} \eqno{(5)}$$
Another method of the analysis, which we employed before, is
the Pad\'e approximation for the ratio $f_{k}$ by

$$ f_{k}= A(1-{1\over{k}}){(1 + {b_{1}\over{k}})\over{(1+{c_{1}\over{
k}})}}\eqno{(6)}$$
\noindent
from which the string susceptibility $\gamma_{string}=2+b_{1}-c_{1}$
is estimated.
The unknown values $A,b_{1} $ and $c_{1}$ are determined from
$f_{k}$,$f_{k-1}$ and $f_{k-2}$ for the fixed order $k$. This Pad\'e
becomes exact in the pure gravity $a=0$ case
with the value $b_{1}=-1/2$
and $c_{1}=2$. This analysis is more convergent and it is convenient
to observe the singularity due to the finite order perturbation,
especially for the separated peaks at the large value of $a$. This
method however fails to obtain the correct result for $n=2 (c=1)$
 since (6) has no $log k$ correction. For $n=2$, we have checked that
our eighth order perturbation agrees with $C_{k}\simeq A^{k}k^{-3}/
log k$ behavior, for $a=1/4$.
 So indeed we have $\gamma_{string}=0$ for $n=2$, although we could
not determine with precision the critical value 1/4 of a. The
1/logk in $C_{k}$ is indeed compatible with the logarithmic deviation to
KPZ scaling found in the c=1 problem.\ref{11}
\par
In Fig.1a-1c, the string susceptibility $\gamma_{string}=2+b_{1}
-c_{1}$ in (6)
is plotted for
n=1, n=4, and n=6  respectively. At $a=0$ and $a=1$, the string
susceptibility exponent becomes exactly $\gamma_{string}=-1/2$. And it
takes the crossover from the pure gravity fixed point to the new
fixed point at the critical vaue of $a$.
 In Fig.1a-1c, the lines of a,b,c and d
correspond to the
Pad\'e approximaion (6) for the string susceptibility exponent
of order $g^{5}$,$g^{6}$,$g^{7}$ and $g^{8}$
respectively. In Fig 1a (n=1),
we find  that at the  critical point $a=1/4$,
$\gamma_{string}=-0.34$, which is very close to the exact value of -1/3.
The critical value of a is known from the exact solution.\ref{7}
  For $n=3$, we have a slightly negative peak value of
$\gamma_{string}$
by this method. There is a logarithmic singularity for n=2,
this may still influence the n=3 series, even if there is no
log singularity in that case.
The ratio metod of (5) for $n=3$ case gives
$\gamma_{string}\simeq 0.02$.
For $n=4$, $\gamma_{string}$ becomes 0.04 as shown in Fig 1b.
The n=2 logarithmic singularity may still lead to
underestimating the exponent for n=4.
  In the case $n=6$ (Fig.1c),
a second peak appears for larger values of a,
but its height decreases at
higher orders, and this second peak seems to merge with the first one
near $a=0.2$.
We estimate the string susceptibility $\gamma_{string}$ as 0.14
for $n$=6. The broken line is a guide line and the crossing point
is the extrapolated value. In $n$=8,
the second peak becomes more pronounced but
it also approaches to the first small peak around $a=0.15$.
In this case,
assuming the two peaks merge at the same point, we estimate
$\gamma_{string}\simeq 0.25$.
The obtained values up to $n$=8 ($c$=4) are not far from
 $(c-1)/12$, which is
the real part of the KPZ formula,\ref{12} valid for $c<1$,

$$ \gamma_{string} = {{c-1}\over{12}}-{\sqrt{(c-1)(c-25)}\over{12}}
\eqno{(7)}$$
\noindent
although our values are lower.
\par
The free energy of the genus one is  evaluated
from the diagram of the
genus zero by permutation of the four lines at the vertices.
We calculate the ratio of each order of genus for a fixed diagram, and then
normalize the coefficients of each diagram by the genus zero results.
 The genus one case for the pure gravity has been studied\ref{13},
from which we  checked our perturbation series. The large order
behavior of the coefficient $C_{k}$ of the genus one of the free energy is
also written by

$$ C_{k} \simeq A^{k}k^{-3+\tilde\gamma_{string}}\eqno{(8)}$$
We have a scaling relation of $\tilde\gamma_{string}$ for $h$-genus case,

$$ \tilde\gamma_{string}= \gamma_{string} + h \gamma_{string}'
 \eqno{(9)}$$
For the genus one, we have $h=1$,
$$    \gamma_{string} + \gamma_{string}' = 2 \eqno{(10)}$$

\noindent
if $c<1$\ref{14}. This relation may be violated for $c>1$ and we are unable to
assume this relation.
 The free energy of
the genus one for the Hermitian matrix model is expanded up to
order six, and here we present the result up to fifth order result,

$$\eqalign{ &E_{1}(g) = -g + g^{2}[ 20(1+a^2)^{n}+10(1+a^{4})^{n}]\cr
             &-g^{3}[256(1+a^{2})^{2n}+416(1+a^{2}+2a^{4})^{n}
           +{608\over{3}}(1+3a^{2})^{n}
           +{544\over{3}}(1+3a^{4})^{n}]\cr
           &+g^{4}[2816(1+a^{2})^{3n}+6400(1+4a^{2}+3a^{4})^{n}+
         2080(1+6a^{2}+a^{4})^{n}\cr
         &+896(1+a^{2}+4a^{4}+a^{6}+a^{8})^{n}+
          880(1+6a^{4}+a^{8})^{n}\cr
         &+9856(1+a^{2}+5a^{4}+a^{6})^{n}+
          6144(1+2a^{2}+5a^{4})^{n}
         +4480(1+3a^{2}+3a^{4}+a^{6})^{n}
         \cr
        &+3712(1+2a^{2}+3a^{4}+2a^{6})^{n}+
         2912(1+5a^{4}+2a^{6})^{n}]\cr
&-g^{5}[28672(1+a^{2})^{4n}+63488((1+a^{2})(1+4a^{2}+3a^{4}))^{n}\cr
&+32768((1+a^{2})(1+2a^{2}+3a^{4}+2a^{6}))^{n}+
78848((1+a^{2})(1+6a^{2}+a^{4}))^{n}\cr
&+77824((1+a^{2})(1+3a^{2}+3a^{4}+a^{6}))^{n}\cr
&+107520((1+a^{2})(1+2a^{2}+5a^{4}))^{n}\cr
&+81920((1+a^{2})(1+a^{2}+5a^{4}+a^{6}))^{n}
+34816(1+3a^{2})^{2n}\cr
&+34304((1+3a^{2})(1+a^{2}+2a^{4}))^{n}
+21606.4(1+10a^{2}+5a^{4})^{n}\cr
&+49664(1+6a^{2}+5a^{4}+4a^{6})^{n}
+125952(1+4a^{2}+9a^{4}+2a^{6})^{n}\cr
&+92672(1+3a^{2}+7a^{4}+5a^{6})^{n}
+47616(1+2a^{2}+7a^{4}+4a^{6}+2a^{8})^{n}\cr
&+18944(1+3a^{2}+6a^{4}+3a^{6}+3a^{8})^{n}
+176128(1+2a^{2}+9a^{4}+4a^{6})^{n}\cr
 &+60928(1+a^{2}+9a^{4}+3a^{6}+2a^{8})^{n}\cr
  &+ 38912(1+3a^{2}+11a^{4}+a^{6})^{n}
   + 97792(1+a^{2}+7a^{4}+7a^{6})^{n}\cr
&+44032(1+2a^{2}+10a^{4}+2a^{6}+a^{8})^{n}
  +102400(1+a^{2}+8a^{4}+5a^{6}+a^{8})^{n}\cr
&+6400(1+2a^{2}+5a^{4}+4a^{6}+4a^{8})^{n}
+29696(1+a^{2}+7a^{4}+3a^{6}+4a^{8})^{n}\cr
&+4710.4(1+10a^{4}+5a^{8})^{n}+
29184(1+8a^{4}+4a^{6}+3a^{8})^{n}\cr
&+46592(1+7a^{4}+6a^{6}+2a^{8})^{n}
  +63488((1+a^{2})(1+4a^{2}+3a^{4}))^{n}\cr
&+3942.4(1+5a^{4}+10a^{6})^{n}]
+ \dots\cr} \eqno{(11)}$$
At $a=0$, we have the correct expression obtained before,\ref{13}
$$\eqalign{ E_{1}&(g)=-g+30g^{2}-1056g^{3}+40176g^{4}
-1600819.2g^{5}\cr
&+65774592g^{6}-2762461769g^{7}+1.17944875\times 10^{11}g^{8}-\dots
\cr}\eqno{(12)}$$
The string susceptibility $\tilde\gamma_{string}$ is obtained by the ratio
method
of (5) as shown in Fig.2a and Fig.2b for $n$=1 and $n$=3 respectively.
Since the successive lines in Fig.2 are approaching smoothly to the line
$\tilde\gamma_{string}$=2 especially for the small value of $a$ ($a
<0.25$), the
results support the  result $\tilde\gamma_{string}=2$. For $n=3$
there is a tiny peak around $a=0.2$. Even if this relation is violated,
the value of $\tilde \gamma_{string}$ remains close to two.
Other cases of different $n$ have similar curve as Fig.2 and
the relation of (10) seem to hold at least when the critical value $a$
remains to be small enough.\par
For $n=2$ case, we find that the coefficient $C_{k}$ in genus one
behaves as (8)
without log $k$ correction. With the assumption of the
log $k$ correction, we are unable to
hold the relation $\tilde \gamma_{string}=2$. This rules out
logarithmic terms at genus one.
\par
We have also developed the expansion for the genus-two case up to
order $g^{6}$. By the simple
ratio method, in which the ratio $C_{k}/C_{k-1}$ is plotted against
$1/k$, we find the same value of the
critical cosmological constant $A$
from three different genus cases. It seems that genus one case gives
most reliable value of the critical cosmological constant $A$. The scaling
relation of (9) seems to hold.
The renormalization group approach for the matrix model has been
investigated.\ref{14} We will discuss this approach for the present
model elsewhere.

The authors thank Al.Zamolodchikov for
his interest and his comments.
\par
\centerline{{\bf References }}
\par
\item{1.} E. Br\'ezin and S. Hikami, Phys. Lett. {\bf B283} (1992)
203.
\item{2.} C. Baillie and D. Johnston, Phys. Lett. {\bf B286 } (1992)
44.
\item{3.} S. M. Catterall, J. B. Kogut and R. L. Renken, a preprint,
    hep-lat.9206021.
\item{4.} J. Ambjorn, B. Durhuus and J. Frohlich, Nucl. Phys. {\bf
B257} (1985) 433.
\item{5.} F. David, Nucl. Phys. {\bf B257} (1985) 543.
\item{6.} D.V. Boulatov, V.A. Kazakov, I. Kostov and A. Migdal,
    Nucl. Phys. {\bf B275} (1986) 641.
\item{7.} D.V. Boulatov and V.A. Kazakov, Phys. Lett. {\bf B214 }
 (1988) 581.
\item{8.} E. Br\'ezin, A. Fujita and S. Hikami, Phys. Rev. Lett.
{\bf 65} (1990) 1949.
\item{9.} M. L. Mehta, Comm. Math. Phys. {\bf 49} (1981) 327.
\item{10.} E. Br\'ezin, C. Itzykson, G. Parisi and J. B. Zuber,
 Comm. Math. Phys. {\bf 59} (1978) 35.
\item{11.} E. Br\'ezin, V. Kazakov and Al. A. Zamolodchikov,
           Nucl. Phys. {\bf B338} (1990) 673.
\item{   }  P. Ginsparg and J. Zinn-Justin, Phys. Lett.{\bf B240}
           (1990) 333.
\item{   }  D. J. Gross and N. Miljkovic, Phys. Lett.{\bf B238} (1990) 217.
\item{   }  G. Parisi, Phys. Lett. {\bf B238} (1990) 209.
\item{12.} V.G.Knizhnik, A.M.Polyakov and A.A.Zamolodchikov,
     Mod. Phys.Lett.{\bf A3} (1988) 819.
\item{   } F.David, Mod. Phys. Lett.{\bf A3} (1988) 207.
\item{   }J. Distler and H.Kawai, Nucl. Phys. {\bf B231} (1989) 509.
\item{13.} D. Bessis, Comm. Math. Phys. {\bf 69} (1979) 147.
\item{14.} E. Brezin and J. Zinn-Justin, Phys. Letters {\bf B288}
         (1992) 54.
\vfill
\eject
\centerline{\bf Figure caption}
\item{Fig.1a} $\gamma_{string}$ for n=1 Ising case. At a=1/4, $\gamma_{string}$
              becomes -0.34. The lines a,b,c and d are Pad\'e analysis
              of (6) from order $g^{5}$,$g^{6}$,$g^{7}$ and $g^{8}$,
              respectively.
\item{Fig.1b} $\gamma_{string}$ for n=4 (c=2). a,b,c and d are same as
              Fig.1a.
\item{Fig.1c} $\gamma_{string}$ for n=6 (c=3). a,b,c and d are same as
              Fig.1a. The broken lines are guided lines for the extrapolation.
\item{Fig.2a.} $\tilde \gamma_{string}$ of n=1 for genus one. The broken line
              is $\tilde \gamma_{string} = 2$. The lines a,b,c and d
              are the result of the ratio method (5) for order $g^{3}$
              ,$g^{4}$,$g^{5}$ and $g^{6}$, respectively.
\item{Fig.2b}  $\tilde \gamma_{string}$ of n=3 for genus one. The lines
              a,b,c and d are same as Fig.2a.
\vfill
\eject

\bye